\documentclass[manuscript]{aastex}

\usepackage{graphicx,times}
\shorttitle{Turbulent convection model in the overshooting region}
\shortauthors{Zhang \& Li}

\begin{document}

\title{Turbulent convection model in the overshooting region: II. Theoretical analysis}
\author{Q.S. Zhang\altaffilmark{1,2,3} and Y. Li\altaffilmark{1,2}}
\email{zqs@ynao.ac.cn(QSZ); ly@ynao.ac.cn(YL)}

\altaffiltext{1}{National Astronomical Observatories/Yunnan Observatory, Chinese Academy of Sciences, P.O. Box 110, Kunming 650011, China.}
\altaffiltext{2}{Laboratory for the Structure and Evolution of Celestial Objects, CAS.}
\altaffiltext{3}{Graduate School of Chinese Academy of Sciences, Beijing 100039, China.}

\begin{abstract}
 Turbulent convection models are thought to be good tools to deal with the convective overshooting in the stellar interior. However, they are too complex to be applied in calculations of stellar structure and evolution. In order to understand the physical processes of the convective overshooting and to simplify the application of turbulent convection models, a semi-analytic solution is necessary.
 We obtain the approximate solution and asymptotic solution of the turbulent convection model in the overshooting region, and find some important properties of the convective overshooting:
 I. The overshooting region can be partitioned into three parts: a thin region just outside the convective boundary with high efficiency of turbulent heat transfer, a power law dissipation region of turbulent kinetic energy in the middle, and a thermal dissipation area with rapidly decreasing turbulent kinetic energy. The decaying indices of the turbulent correlations $k$, $\overline{u_{r}'T'}$, and $\overline{T'T'}$ are only determined by the parameters of the TCM, and there is an equilibrium value of the anisotropic degree $\omega$.
 II. The overshooting length of the turbulent heat flux $\overline{u_{r}'T'}$ is about $1H_k$($H_k=|\frac{dr}{dlnk}|$).
 III. The value of the turbulent kinetic energy at the convective boundary $k_C$ can be estimated by a method called \textsl{the maximum of diffusion}. Turbulent correlations in the overshooting region can be estimated by using $k_C$ and exponentially decreasing functions with the decaying indices.

\end{abstract}


\section{Introduction}

Convective overshooting is an important physical process in the stellar structure and evolution.
 Phenomenologically, the acceleration of a fluid element is zero at the convective boundary,
 but its speed is not zero. It is able to go across the convective boundary into the dynamically stable zone. This phenomenon is called the convective overshooting.
 The convective overshooting transports heat and matter, and affects the structure and evolution of stars. A phenomenological theory of the overshooting was developed by Zahn(1991), which predicts an adiabatic overshooting region. However, Xiong \& Deng(2001) pointed out that the turbulent velocity and the temperature are strongly correlated in Zahn's theory. Recently, Christensen-Dalsgaard et al.(2011) found that the convective overshooting only described by the turbulent convection models could be in agreement with the helioseismic data.

The turbulent convection models (TCMs) are based on fully
hydrodynamic moment equations,
 and applied on investigating the convective overshooting\citep{xio81,xio85,xio89,xio01,can97,can98,can98b,can99,mar02,den06,li07,den08,zha09}. There are two main difficulties restricting the applications of the TCMs. One is to solve the equations of the TCMs, which are highly non-linear and unstable in numerical calculations. The other is to incorporate the TCMs into a stellar evolution code. In general, solving the TCMs needs the parameters of the stellar structure(e.g. temperature $T$, density $\rho$, pressure $P$, radius $r$, luminosity $L$, and elements abundance vector), and solving the equations of stellar structure requires the temperature gradient $\nabla$ which is determined by the TCMs. Thus, in order to apply the TCMs, one must solve both the TCMs and the equations of stellar structure, which shows enormous difficulty. Although developing numerical technique is very important, getting an approximate solution of the TCMs is more interesting because an approximate solution helps to understand the physical processes and may significantly simplify the application of the TCMs. Xiong(1989) found the asymptotic solution of his TCM in the overshooting region, the turbulent correlations being exponentially decreasing in the overshooting region. However, his solution of the heat flux $\overline{u_{r}'T'}$ is not suitable near the convective boundary, and the initial turbulent kinetic energy $k_0$ is unknown so that the value of the turbulent correlations in the overshooting region actually can not be determined without numerical calculations.

In this paper, we investigate the properties of the convective overshooting by analyzing Li \& Yang's TCM\citep{li07}, which was tested in the solar convection zone\citep{li07,yan07}. We try to get a semi-analytical solution of the TCM in the overshooting region. We introduce the TCM in Section 2, investigate the properties of the overshooting in Section 3, and summarize the conclusions in Section 4.

\section{Turbulent Convection Model}

The closure assumptions of Li \& Yang's TCM are\citep{li01,li07}: the three-order moment terms are modeled with a gradient-type scheme; the dissipation rate $\varepsilon$ of the turbulent kinetic energy $k$ is assumed to be local; the dissipation rates of the turbulent heat flux $\overline{u_{r}'T'}$ and the turbulent fluctuation of temperature $\overline{T'T'}$ are assumed to be determined by both the reciprocal timescale of the turbulent dissipation $\tau_1^{-1}=\frac{\varepsilon}{k}$ and the thermal dissipation one $\tau_2^{-1}=\frac{\lambda }{\rho c_{P}}\frac{\varepsilon ^{2}}{k^{3}}$. According to those closure assumptions, fully hydrodynamic moment equations on the quasi-steady approximation result in the complete equations of two-order moment terms\citep{li07}:
\begin{eqnarray}
\frac{1}{\rho r^{2}}\frac{\partial }{\partial r}\left( C_{s}\rho r^{2}\frac{k%
}{\varepsilon }\overline{u_{r}'u_{r}'}\frac{\partial
\overline{u_{r}'u_{r}'}}{\partial r}\right)
 =\frac{2}{3}%
\varepsilon +\frac{2\beta g_{r}}{T}\overline{u_{r}'T'}
+C_{k}\frac{\varepsilon }{k}\left( \overline{u_{r}'u_{r}'}-%
\frac{2}{3}k\right)
\end{eqnarray}%
\begin{eqnarray}
\frac{1}{\rho r^{2}}\frac{\partial }{\partial r}\left( C_{s}\rho r^{2}\frac{k%
}{\varepsilon }\overline{u_{r}'u_{r}'}\frac{\partial k}{%
\partial r}\right) =\varepsilon +\frac{\beta g_{r}}{T}\overline{%
u_{r}'T'}
\end{eqnarray}%
\begin{eqnarray}
\frac{2}{\rho r^{2}}\frac{\partial }{\partial r}\left( C_{t1}\rho r^{2}\frac{%
k}{\varepsilon }\overline{u_{r}'u_{r}'}\frac{\partial
\overline{u_{r}'T'}}{\partial r}\right)
=-\frac{T}{H_{P}}%
(\nabla-\nabla_{ad})\overline{u_{r}'u_{r}'}
+\frac{\beta
g_{r}}{T}\overline{T'T'}+C_{t}\left( \frac{\varepsilon }{k}+\frac{\lambda
}{\rho c_{P}}\frac{\varepsilon ^{2}}{k^{3}}\right)
\overline{u_{r}'T'}
\end{eqnarray}%
\begin{eqnarray}
\frac{1}{\rho r^{2}}\frac{\partial }{\partial r}\left( C_{e1}\rho r^{2}\frac{%
k}{\varepsilon }\overline{u_{r}'u_{r}'}\frac{\partial \overline{T'T'}}{\partial r}\right)
=-\frac{2T}{H_{P}}%
(\nabla-\nabla_{ad})\overline{u_{r}'T'}
+2C_{e}\left(
\frac{\varepsilon }{k}+\frac{\lambda }{\rho c_{P}}\frac{%
\varepsilon ^{2}}{k^{3}}\right) \overline{T'T'}
\end{eqnarray}%

The temperature gradient is calculated as:
\begin{eqnarray}
\nabla=\nabla_{R}-\frac{H_P}{T}\frac{\rho c_{P}}{\lambda}\overline{u_{r}'T'}
\end{eqnarray}%

The meaning of those equations and each term in them were described in previous works\citep{li07,zha09} in detail.
We simply introduce them here:

Equations (1-4) describe the equilibrium(time-independent) structure of the radial kinetic energy $\overline{u_{r}'u_{r}'}$, the turbulent kinetic energy $k$, the turbulent heat flux $\overline{u_{r}'T'}$ and the turbulent fluctuation of temperature $\overline{T'T'}$, respectively. On the left side of those equations, there is the non-local term(i.e. the diffusion term) of each turbulent correlation. On the right side, there are the local terms which describe the generation and the dissipation of each turbulent correlation.

In Eq.(1) and (2), $\varepsilon$ is the turbulent dissipation rate of $k$ and $\varepsilon=\frac{k^\frac{3}{2}}{l}$ where $l=\alpha H_P$, and the second term on the right side is the generation rate of the kinetic energy due to the contribution of the buoyancy. The last term in Eq.(1) is the \textsl{return to isotropy} term which attempts to make the turbulent motion be isotropic.
In Eq.(3), the first two terms on the right side is the generation rate of the turbulent heat flux $\overline{u_{r}'T'}$, and the last one is the dissipation rate that comprises the turbulent dissipation and the thermal dissipation.
In Eq.(4), the first term on the right side is the generation rate of the turbulent fluctuation of temperature $\overline{T'T'}$, and the last one is the dissipation rate.
Meanings of other symbols are: $H_P=-\frac{dr}{dlnP}$ is the local pressure scale height, $\beta=-(\frac{\partial ln \rho}{\partial ln T})_P$ the expansion coefficient, $g_r = - \frac{GM_r}{r^2}$ the radial component of gravity acceleration, $\nabla=\frac{dlnT}{dlnP}$ the temperature gradient in the stellar interior, $\nabla_{ad}=(\frac{\partial lnT}{\partial lnP})_S$ the adiabatic temperature gradient, $\lambda=\frac{4acT^3}{3\kappa\rho}$ the thermal conduction coefficient, $c_P=(\frac{\partial H}{\partial T})_P$ the specific heat, $C_k$ the parameter of the \textsl{return to isotropy} term, ($C_s,C_{t1},C_{e1}$) the diffusion parameters and ($\alpha,C_t,C_e$) the dissipation parameters of turbulent variations($k,\overline{u_{r}'T'},\overline{T'T'}$).

In Eqs.(1-4), overbars are only used in three turbulent correlations $\overline{u_{r}'u_{r}'}$, $\overline{u_{r}'T'}$ and $\overline{T'T'}$. The other variations(density $\rho$ and the temperature $T$, etc.) are all mean state quantities which should use overbars but we ignore them for convenience.

Equation (5) describes the energy transport in the stellar interior by both turbulent motions(i.e. convection and overshooting) and radiation. $\nabla_R$ is the radiative temperature gradient.

\section{Theoretical analysis of TCM in the overshooting region}

In the previous work \citep{zha09},
we applied the TCM in the solar overshooting region and found some properties of the overshooting region:
$\overline{u_{r}'T'}<0$, $\nabla_{R}<\nabla<\nabla_{ad}$, and the peak of $\overline{T'T'}$, which are similar to Xiong's(1985) and Xiong \& Deng's(2001) works. In this section, we attempt to get semi-analytical solutions of the TCM.

Some approximations are adopted to simplify Eqs.(1-5) in the overshooting region:

\textsl{Approximation I}. P\'{e}clet number $P_e \gg 1$, where $P_e = \frac{ \rho C_P l \sqrt{k}}{\lambda}$. That is $\frac{\varepsilon }{k}\gg\frac{\lambda }{\rho c_{P}}\frac{%
\varepsilon ^{2}}{k^{3}}$ which means the turbulent dissipation is much stronger than the thermal dissipation. This assumption is reasonable in most cases except for the region near the surface of a star or with very small $k$.

\textsl{Approximation II}. All variations, except the turbulent fluctuations, are thought to be constant because the turbulent fluctuations
change much faster than others in the overshooting region.

\textsl{Approximation III}. Far away from the convective boundary, $\nabla\approx\nabla_{R}$. This assumption is acceptable if the heat flux $\overline{u_{r}'T'}$ is small.

\subsection{Turbulent heat transport in the overshooting region}

Defining the anisotropic degree $\omega=\frac{\overline{u_{r}'u_{r}^{\prime
}}}{2k}$ which is the ratio of radial kinetic energy to total kinetic energy, and applying \textsl{Approximation II} and Eq.(5), we can rewrite Eq.(3) to:
\begin{eqnarray}
\frac{\partial }{\partial r}\left( 4C_{t1}  \omega l\sqrt{k}\frac{\partial
\overline{u_{r}'T'}}{\partial r}\right) =-\frac{T}{H_{P}}%
(\nabla_R-\nabla_{ad})\overline{u_{r}^{\prime}u_{r}'}
+\frac{\beta g_{r}}{T}\overline{T'T'}+[2\omega P_e+C_t (1+P_e^{-1})]
\frac{\sqrt{k}}{l}\overline{u_{r}'T'}
\end{eqnarray}%

In the last bracket in Eq.(6), \textsl{Approximation I}($P_e \gg 1$) makes the dissipation term $C_t (1+P_e^{-1})\frac{\sqrt{k}}{l}\overline{u_{r}'T'}$ be ignorable. And, by using Eq.(5) and \textsl{Approximation II}, it is easy to find that the diffusion term is on the same order of the ignorable dissipation term:
\begin{eqnarray}
\frac{\partial }{\partial r}\left( 4C_{t1}  \omega l\sqrt{k}\frac{\partial
\overline{u_{r}'T'}}{\partial r}\right) \approx 2C_{t1}\alpha^2\omega
\frac{dlnk}{dlnP}\cdot\frac{dln(\nabla_R-\nabla)}{dlnP} (\frac{\sqrt{k}}{l}\overline{u_{r}'T'})
\sim Pe^0 (\frac{\sqrt{k}}{l}\overline{u_{r}'T'})
\end{eqnarray}%

Therefore the diffusion term is also ignorable.
Equation (3) is in local equilibrium:
\begin{eqnarray}
-\frac{T}{H_{P}}%
(\nabla-\nabla_{ad})\overline{u_{r}^{\prime
}u_{r}'}+\frac{\beta
g_{r}}{T}\overline{T'T'}\approx0
\end{eqnarray}%

In the overshooting region, the most important process is the diffusion of the kinetic energy. Thus, we ignore the diffusion of $\overline{T'T'}$(i.e., setting $C_{e1}=0$). The solution of the TCM with $C_{e1}=0$ can be thought as the zero-order solution of the TCM.

Ignoring the diffusion of $\overline{T'T'}$ and the diffusion and dissipation terms of $\overline{u_r'T'}$, using \textsl{Approximations I \& II}, one can rewrite Eqs.(1-4) as:
\begin{eqnarray}
\frac{2C_{s}l}{k}\frac{\partial }{\partial r}( \omega k^{\frac{5}{2}}\frac{\partial \omega}{\partial r}) =(C_k-1)(\omega-\frac{1}{3})\frac{k^{\frac{3}{2}} }{l}
 +\frac{\beta g_{r}}{T}\overline{u_{r}'T'}(1-\omega)
\end{eqnarray}%
\begin{eqnarray}
2C_{s}l\frac{\partial }{\partial r}( \omega k^{\frac{1}{2}}\frac{\partial k}{\partial r}) =\frac{k^{\frac{3}{2}} }{l}
 +\frac{\beta g_{r}}{T}\overline{u_{r}'T'}
\end{eqnarray}%
\begin{eqnarray}
0=-\frac{2T}{H_{P}}
(\nabla-\nabla_{ad})\omega k+\frac{\beta g_{r}}{T}\overline{T'T'}
\end{eqnarray}%
\begin{eqnarray}
0=-\frac{2T}{H_{P}}
(\nabla-\nabla_{ad})\overline{u_{r}'T'}+2C_{e} \frac{\varepsilon }{k} \overline{T'T'}
\end{eqnarray}%

Equation (9) results from Eq.(1) and (2), describing the equilibrium structure of the anisotropic degree $\omega$. The left side is the diffusion of $\omega$. The first term in the right side is the dissipation rate due to \textsl{return to isotropy} term in Eq.(1). The last term is the generation rate of $\omega$ due to the buoyancy.

Equations (11) and (12) show:
\begin{eqnarray}
0=(\nabla-\nabla_{ad})(\overline{u_{r}'T'}+ 2C_{e} \omega\frac{T}{\beta g}\frac{k^{\frac{3}{2}} }{l})
\end{eqnarray}%

The solution is $\overline{u_{r}'T'}=- 2C_{e}
\varepsilon\omega\frac{T}{\beta g}$ or $\nabla=\nabla_{ad}$. The latter is equivalent to $\overline{u_{r}'T'}=-\frac{T}{H_P}\frac{\lambda}{\rho c_{P}}(\nabla_{ad}-\nabla_R)$. Because $\overline{u_{r}'T'}$ is close to zero near the convective boundary and gradually decreases far away from the convective boundary\citep{xio89,xio01,zha09}, the physically acceptable result is:
\begin{eqnarray}
\overline{u_{r}'T'}=Max\{-\frac{T}{H_P}\frac{\lambda}{\rho c_{P}}(\nabla_{ad}-\nabla_R),
- 2C_{e}\omega\frac{T}{\beta g}\frac{k^{\frac{3}{2}} }{l}\}
\end{eqnarray}%

Equation (14) shows that there is an adiabatic stratification zone in the overshooting region in the case of $C_{e1}=0$. In order to investigate the property of heat transport in the overshooting region, we must know the length of the adiabatic stratification zone. It is found in Eq.(14) that the boundary of the adiabatic stratification is the location where $\frac{T}{H_P}\frac{\lambda}{\rho c_{P}}(\nabla_{ad}-\nabla_R)=2C_{e}
\omega\frac{T}{\beta g}\frac{k^{\frac{3}{2}} }{l}$. Solving the equation of $\omega$ is not easy because it is nonlinear. However, this problem is avoidable. Turbulent motions are isotropic when $\omega=\frac{1}{3}$. In the convection zone, $\omega>\frac{1}{3}$ because the buoyancy boosts radial turbulent motion. In most part of overshooting region, $\omega$ should be less than $\frac{1}{3}$ because the buoyancy prevents radial turbulent motion. Therefore $\omega$ should be not far away from $\frac{1}{3}$  near the convective boundary. Further more, taking $\omega$ as a constant, one can rewrite Eq.(10) as:
\begin{eqnarray}
2C_{s}l\omega \frac{\partial }{\partial r}( k^{\frac{1}{2}}\frac{\partial k}{\partial r})
=\frac{k^{\frac{3}{2}} }{l} +\frac{\beta g_{r}}{T}\overline{u_{r}'T'}
\end{eqnarray}%

Substituting Eq.(14) into the above equation, one can get the approximate solution:
\begin{eqnarray}
k^{\frac{3}{2}}\approx k_C^{\frac{3}{2}}exp(-\sqrt{\frac{3}{4C_s \omega}} |\frac{r-r_C}{l}|)
\end{eqnarray}%
if $\frac{T}{H_P}\frac{\lambda}{\rho c_{P}}(\nabla_{ad}-\nabla_R) \leq 2C_{e}
\omega\frac{T}{\beta g}\frac{k^{\frac{3}{2}} }{l}$, and:
\begin{eqnarray}
k^{\frac{3}{2}}=k_A^{\frac{3}{2}}exp(-\sqrt{\frac{3(1+2C_e \omega)}{4C_s \omega}} |\frac{r-r_A}{l}|)
\end{eqnarray}%
if $\frac{T}{H_P}\frac{\lambda}{\rho c_{P}}(\nabla_{ad}-\nabla_R) > 2C_{e}
\omega\frac{T}{\beta g}\frac{k^{\frac{3}{2}} }{l}$.

In Eq.(16), point C, which is the convective boundary where $\nabla_{ad}=\nabla_R$, is set to be the initial point, $k_C$ and $r_C$ being $k$ and $r$ here. The contribution of the buoyancy term(i.e. the last term in Eq.(15)) is ignored in obtaining the solution Eq.(16). In the deep convection zone, turbulent motions are almost in local equilibrium, thus the ratio of $-\frac{\beta g_{r}}{T}\overline{u_{r}'T'}$ to $\frac{k^{\frac{3}{2}} }{l}$ is about 1. However, near the convective boundary, buoyancy is about zero, meanwhile the diffusion of $k$ dominates. Those make the ratio be much less than 1. Therefore the buoyancy term is ignorable.

In Eq.(17), point A, where $k=k_A$ and $r=r_A$, is the boundary of the adiabatic overshooting region. In the region beyond point A, the ratio of $-\frac{\beta g_{r}}{T}\overline{u_{r}'T'}$ to $\frac{k^{\frac{3}{2}} }{l}$ is $2C_e\omega$ which is on the order of $1$, thus the buoyancy term remains.

The exponentially decreasing function of $k$ is due to the fact that there is no generation in the
overshooting region. Contrary to the situation in the convection zone, the buoyancy dissipates $k$ because it
prevents the radial motion of fluid elements in the overshooting region. The distribution of $k$ results from the equilibrium between the diffusion and the dissipation.
$k$ should decrease faster if the buoyancy is as effective as the turbulent
dissipation, which is found by comparing the exponential indices of
Eq.(16) and (17).

The location of point A is determined by $\frac{T}{H_P}\frac{\lambda}{\rho c_{P}}(\nabla_{ad}-\nabla_R) = 2C_{e}
\omega\frac{T}{\beta g}\frac{k^{\frac{3}{2}} }{l}$. Using Eq.(16), we get a property of point A:
\begin{eqnarray}
k_C^{\frac{3}{2}}exp(-\sqrt{\frac{3}{4C_s \omega}} |\frac{r_A-r_C}{l}|)
=\frac{1}{2C_{e}\omega} \frac{\alpha\beta g\lambda}{\rho c_{P}}(\nabla_{ad}-\nabla_{R,A})
\end{eqnarray}%

The relation between $r_A$ and $\nabla_{R,A}$ is needed in order to solve this equation and to locate point A.
Near the convective boundary, there is:
\begin{eqnarray}
|\nabla_{ad}-\nabla_{R,A}|\approx \nabla_{ad}|\chi(lnP_A-lnP_C)|=\nabla_{ad}|\chi|\frac{l_{ad}}{H_P}
\end{eqnarray}%
where $l_{ad}=|r_A-r_C|$ is the length of the adiabatic overshooting region, $P_A$ and $P_C$ the pressure at point A and C, and $\chi=\frac{dln\nabla_{R}}{dlnP}$ which is approximately a constant.

Substituting Eq.(19) into Eq.(18), one finds:
\begin{eqnarray}
k_C^{\frac{3}{2}}exp(-\frac{1}{\alpha}\sqrt{\frac{3}{4C_s \omega}} \frac{l_{ad}}{H_P})
=\frac{1}{2C_{e}\omega} \frac{\alpha\beta g\lambda}{\rho c_{P}}\nabla_{ad}|\chi|\frac{l_{ad}}{H_P}
\end{eqnarray}%

$l_{ad}$ can be worked out if $k_C$ is known. In the deep adiabatic convection zone, turbulent diffusion is ignorable, and the localized TCM shows $k^{\frac{3}{2}}_{Local}=\frac{ \alpha\beta g\lambda(\nabla_{R}-\nabla_{ad})}{\rho c_{P}}$ (see Appendix A). However, $k_C$ can not be estimated as that because $\nabla_{R}=\nabla_{ad}$ thus $k_{Local}=0$ at the convective boundary. Actually, the turbulent diffusion of $k$ is effective near the convective boundary, and $k_C$ is determined by the diffusion. We can estimate $k_C$ by a simple approach which will be referred to as \textsl{the maximum of diffusion} hereafter. Setting point B at where the diffusion becomes dominative in the convection zone, we get the relation between $k_C$ and $k_B$ by solving Eq.(15):
\begin{eqnarray}
k_C^{\frac{3}{2}}=k_B^{\frac{3}{2}}exp(-\sqrt{\frac{3}{4C_s \omega}} |\frac{r_C-r_B}{l}|)
\end{eqnarray}%
where $k_B^{\frac{3}{2}}\approx \frac{ \alpha\beta g\lambda(\nabla_{R,B}-\nabla_{ad})}{\rho c_{P}}$. Equation (21) shows that $k_C$ is a function of $r_B$. In reality, the diffusion leads to the maximum of $k_C$. Therefore $r_B$ makes the derivation of the right side of Equation (21) be zero. Noting that $\nabla_{R,B}-\nabla_{ad}$ is approximately proportional to $r_B-r_C$, one can easily work out the location of point B:
\begin{eqnarray}
\sqrt{\frac{3}{4C_s \omega}} |\frac{r_C-r_B}{l}|\approx1
\end{eqnarray}%

It is found in Fig.1 that $k\approx k_{Local}$ in the deep convection zone because the turbulent diffusion can be ignored here, and the turbulent diffusion dominates in the layer beyond point B.

Using above results, we obtain:
\begin{eqnarray}
k_C^{\frac{3}{2}}=\frac{1}{e}\frac{ \alpha\beta g\lambda(\nabla_{R,B}-\nabla_{ad})}{\rho c_{P}}
\approx\frac{1}{e}\sqrt{\frac{4C_s \omega}{3}}\frac{ \alpha^2\beta g\lambda\nabla_{ad}|\chi|}{\rho c_{P}}
\end{eqnarray}%

Generally, $\frac{l_{ad}}{H_P}$ is very small. According to Eq.(20), the length of the adiabatic overshooting region is:
\begin{eqnarray}
l_{ad}\approx \frac{\sqrt{\frac{4C_s \omega}{3}}}{\frac{e}{2C_{e}\omega}+1}l
\end{eqnarray}%

In the area $|r-r_C|\leq|r_A-r_C|$ in the overshooting region, the temperature gradient $\nabla$ is almost equal to the adiabatic one. In the area $|r-r_C|>|r_A-r_C|$, however, according to Eq.(14), Eq.(17), and Eq.(5), the temperature gradient $\nabla$ is gradually close to $\nabla_R$:
\begin{eqnarray}
\nabla-\nabla_R=(\nabla_{ad}-\nabla_{R,A})
\cdot exp [-\sqrt{\frac{3(1+2C_e \omega)}{4C_s \omega}} |\frac{r-r_A}{l}|]
\end{eqnarray}%

Although $\omega$ in Eq.(24) and Eq.(25) is still unknown, we can estimate it roughly. Equation (24) and (25) describe the turbulent motion near the convective boundary, thus we can use $\omega\approx\omega_C$ where $\omega_C$ is $\omega$ at the convective boundary. In the deep convection zone, $\omega$ is almost equal to the equilibrium value $\omega_{cz}=\frac{2}{3C_k}+\frac{1}{3}$ which is derived from the localized TCM (see Appendix A). $\omega_C<\omega_{cz}$ because the buoyancy is zero at the boundary, and $\omega_C>\frac{1}{3}$ because the diffusion of $\omega$. Therefore the typical value of $\omega_C$ can be taken as the average, i.e. $\omega_C\approx\frac{1}{2}(\omega_{cz}+\frac{1}{3})$. If Eq.(25) is used in the region far away from the convective boundary(beyond the peak of $\overline{T'T'}$), $\omega\approx\omega_C$ is not appropriate. One can use $\omega=\omega_o$, where $\omega_o$ is the equilibrium value of $\omega$ in the overshooting region which is introduced in the next subsection.

Another turbulent correlation is $\overline{T'T'}$, which can be worked out by using Eq.(11):
\begin{eqnarray}
\overline{T'T'} \approx 0, (|r-r_C|\leq|r_A-r_C|)
\end{eqnarray}%

And:
\begin{eqnarray}
\overline{T'T'}=\frac{2T}{H_{P}}\frac{T}{\beta g}(\nabla_{ad}-\nabla)\omega k, (|r-r_C|>|r_A-r_C|)
\end{eqnarray}%

Equation (26) seems to against Cauchy's theorem $\overline{u_{r}'u_{r}'}\overline{T'T'} \geq \overline{u_{r}'T'}^2$. Actually, ${\overline{T'T'}\approx0}$ is only an approximate solution on the order of ($Pe^1\frac{\sqrt{k}}{l}\overline{u_{r}'T'}$), because Eq.(8) is an approximation on that order. Numerical calculations show no confliction.

Results obtained above are based on $C_{e1}=0$. Numerical results of $\nabla$ with both $C_{e1}=0$ and $C_{e1}\neq0$ are shown in Fig.2. It is found that the effects of the diffusion of $\overline{T'T'}$ are only making $\nabla$ be smoother.
However, there is no adiabatic overshooting region when the diffusion of $\overline{T'T'}$ is present, because $\overline{T'T'}$ increases near the convective boundary due to the turbulent diffusion thus $\nabla$ decreases according to Eq.(8).
Numerical results of the turbulent correlations in both $C_{e1}=0$ and $C_{e1}\neq0$ with different TCM parameters and for different stellar models are shown in Figs.3-5. It is found that the theoretical solutions well fit the numerical solutions in the case of $C_{e1}=0$. This also validates that the boundary value $k_C$ derived from \textsl{the maximum of diffusion} is a good approximation. The diffusion of $\overline{T'T'}$ modifies and smoothes the profile of $\overline{T'T'}$ and $\overline{u_r'T'}$. However, $k$ is insensitive to the diffusion of $\overline{T'T'}$ because that $k$ is mainly dominated by the diffusion of itself. The diffusion of $\overline{T'T'}$ doesn't significantly change the integral value of $\overline{T'T'}$. According to Eq.(8), the integral value of $\nabla$ or $\overline{u_r'T'}$ is also insensitive to the diffusion of $\overline{T'T'}$, which is found in Figs.(2-5).

The distribution of $\overline{T'T'}$ reveals an important property of the overshooting. In the nonadiabatic overshooting region, using $\nabla\approx\nabla_R$, one finds that $\overline{T'T'}\propto T(\nabla_{ad}-\nabla_{R})k$ according to Eq.(27). This result indicates a maximum of $\overline{T'T'}$\citep{xio85,zha09} which is shown in Figs.3-5. Beyond the location of the maximum of $\overline{T'T'}$, the temperature of a turbulent element is gradually close to the temperature of the environment, and the efficiency of heat transport significantly decreases. Therefore the area between the convective boundary and the location of the maximum of $\overline{T'T'}$ can be thought as the overshooting region of $\overline{u_{r}'T'}$. It is found in Figs.3-5 that the width of the valley of $\overline{u_{r}'T'}$ is approximately equal to the distance from the convective boundary to the location of the maximum of $\overline{T'T'}$. In order to get the overshooting length of heat transport, we need to locate the maximum of $\overline{T'T'}$.

Using Eq.(17), defining $\theta_0=\frac{dlnk}{dlnP}=\pm\frac{1}{\alpha}\sqrt{\frac{(1+2C_e \omega)}{3C_s \omega}}$ as the decaying index of $k$ (in the case of $C_{e1}=0$), we get:
\begin{eqnarray}
\overline{T'T'}\propto T(\nabla_{ad}-\nabla_{R})P^{\theta_0}
\end{eqnarray}%

The derivative of $\overline{T'T'}$ is zero at the peak of $\overline{T'T'}$.
We get $\nabla_{R}$ there(denoted as $\nabla_{R}^*$):
\begin{eqnarray}
 (\nabla_R^*+\theta_0)(\nabla_{ad}-\nabla_{R}^*)-\chi\nabla_{R}^* \approx 0
\end{eqnarray}%

$\nabla_{R}^*$ is determined by only one turbulent parameter $\theta_0$.

The typical overshooting length of $\overline{u_{r}'T'}$ (or $\nabla$) can be estimated with $\nabla_{R}^*$:
\begin{eqnarray}
|\chi|=|\frac{dln\nabla_{R}}{dlnP}|\approx\ | \frac{ln\nabla_{R,C}-ln\nabla_{R}^*}{lnP_C-lnP^*}|
=| \frac{ln\nabla_{ad}-ln\nabla_{R}^*}{lnP_C-lnP^*}|=\frac{ln\frac{\nabla_{ad}}{\nabla_{R}^*}}{\frac{l_\nabla}{H_P}}
\end{eqnarray}%
where $\nabla_{R,C}$ is $\nabla_R$ at the convective boundary, $l_\nabla$ is the distance from the convective boundary to the location of the maximum of $\overline{T'T'}$ and also the typical overshooting length of $\nabla$.

$l_\nabla$ is worked out as:
\begin{eqnarray}
l_\nabla\approx\frac{1}{|\chi|}ln\frac{\nabla_{ad}}{\nabla_{R}^*}H_P
\end{eqnarray}%

Usually, $|\theta_0|$ is much larger than $|\chi|$ and $\nabla_{ad}$, and $\nabla_{R}^*$ can be approximately solved from Eq.(29):
\begin{eqnarray}
\nabla_{R}^*\approx(1-\frac{\chi}{\theta_0})\nabla_{ad}
\end{eqnarray}%

Finally, we find:
\begin{eqnarray}
l_\nabla\approx\frac{H_P}{|\theta_0|}=H_k
\end{eqnarray}%
where $H_k$ is the scale height of turbulent kinetic energy $k$ defined by $H_k=|\frac{dr}{dlnk}|$. The result indicates that $\nabla$ is remarkably modified by the overshooting only in about $1H_k$. It is found in Fig.3 that $l_\nabla=ln\frac{k_C}{k_*}H_k \approx 0.8H_k$, which is in agreement with Eq.(33). It is shown in Fig.2 that $\nabla$ is remarkably modified only in $1H_k$.

\subsection{Asymptotic analysis}

In above subsection, we have discussed the turbulent heat transport and the solution of turbulent correlations in the overshooting region near the convective boundary based on the assumption $C_{e1}=0$. The diffusion of $\overline{T'T'}$ only modifies turbulent correlations to be smoother near the convective boundary. However, it makes more effects on turbulent motions in the overshooting region further than $1H_k$ away from the convective boundary. In this subsection, we investigate the turbulence properties in the outer overshooting region(beyond $1H_k$).

In the numerical calculations of the TCM, we found that the anisotropic degree $\omega$ always showed an equilibrium value in the overshooting region. A typical numerical result is shown in Fig.6. In order to understand it, we discuss the behave of the anisotropic degree $\omega$ in both convection zone and overshooting region. $\omega$ should be larger than $\frac{1}{3}$ in the convection zone because the buoyancy boosts radial movement of turbulent elements. Actually, $\omega$ is almost equal to the equilibrium value  in the convection zone $\omega_{cz} = \frac{2}{3C_k}+\frac{1}{3}$(see Appendix A) due to the equilibrium between the buoyancy and the \textsl{return to isotropy} term. When turbulent elements go across the convective boundary into the overshooting region, the buoyancy prevents convective elements moving, thus $\omega$ decreases to less than $\frac{1}{3}$ near the convective boundary. However, as $\overline{u_{r}'T'}$ exponentially decreasing, the equilibrium of $\omega$ is established again in the overshooting region. This results in an asymptotic property of the overshooting region: there is an equilibrium value of $\omega$ in the overshooting region, $\omega\approx\omega_o$.

By using the asymptotic property $\omega\approx\omega_o$ and \textsl{Approximations I, II \& III}, it is easy to get the asymptotic solution of TCM in the overshooting region(see Appendix B):
\begin{eqnarray}
\overline{u_{r}'T'}=\frac{(C_k-1)(\omega_o-\frac{1}{3})}{(1-\omega_o)}\frac{T}{\beta g }\frac{k^{\frac{3}{2}}}{l}
\end{eqnarray}%
\begin{eqnarray}
\overline{T'T'}=2\omega_o(\nabla_{ad}-\nabla_{R})\frac{T^2}{\beta gH_P} k
\end{eqnarray}%
\begin{eqnarray}
k=k_0 (\frac{P}{P_0})^\theta
\end{eqnarray}%
where $\theta$ is the asymptotic solution of $\frac{d ln k}{d ln P}$:
\begin{eqnarray}
\theta=\pm \frac{1}{\alpha}\sqrt{\frac{1}{3C_s\omega_o}[1-\frac{(C_k-1)(\omega_o-\frac{1}{3})}{(1-\omega_o)}]}
\end{eqnarray}%
$k$ takes the decreasing expression in the overshooting region, which means: $'+'$ is adopted in the upward overshooting region and $'-'$ in the downward one.

The equilibrium value $\omega_o$ is determined by:
\begin{eqnarray}
(2C_s C_e-C_{e1}C_k){\omega_o}^2
+[\frac{1}{3}C_{e1}(C_k+2)-C_s(C_k+2C_e-1)]\omega_o
+\frac{1}{3}C_s(C_k-1)=0
\end{eqnarray}%

The equilibrium value $\omega_o$ is only a function of turbulent parameters $(C_e, C_{e1}, C_s, C_k)$. The fact that the buoyancy prevents the radial movement of turbulent elements in the overshooting region restricts the turbulent parameters to ensure $\omega_o<\frac{1}{3}$.

An important thing is where $\omega$ reaches its equilibrium value $\omega_o$. According to Eq.(9), the equilibrium of $\omega$ can be realized only if the buoyancy term synchronically decreases with $k$ decreasing. Therefore $\omega$ starts to reach its equilibrium value $\omega_o$ beyond the peak of $\overline{T'T'}$ due to $|\overline{u_r'T'}|$ being decreasing.

Setting $C_{e1}=0$ in Eq.(38), we find that the asymptotic solution is the same as the results in the overshooting region with $|r-r_C|\geq|r_A-r_C|$ by setting $\omega=\omega_o$ in Eq(14),(17) \& (27). Because Eq.(8) is correct whether $C_{e1}=0$ or not, the conclusion that the maximum of $\overline{T'T'}$ is located at about $1H_k$ is also correct in both cases.

It must be mentioned that we have used \textsl{Approximation I}(i.e. $P_e\gg 1$), which means that the turbulent dissipation is much larger than the thermal dissipation. If $k$ decreases enough to satisfy $P_e\ll 1$, the thermal dissipation should become significant thus $\overline{T'T'}$ and the turbulent kinetic energy $k$ should rapidly decrease to zero. Then $\omega$ also rapidly decreases as shown in Fig.6. In another word, turbulent movement can hardly overshoot into the thermal dissipation zone where $P_e\ll 1$.

According to discussions above, we can separate the overshooting region into three parts as shown in Fig.7: the overshooting region of $\overline{u_r'T'}$ or $\nabla$ with the length of about $1H_k$, the turbulent dissipation region in which the asymptotic solution holds, and the thermal dissipation region in which the turbulent movement quickly vanishes. The boundaries among those parts are the peak of $\overline{T'T'}$ and the location of $P_e=1$.

\section{Conclusions and discussions}

Turbulent convection models are better tools in dealing with the convective overshooting than non-local mixing length theories. However, they are often too complex to be applied in the calculations of stellar structure and evolution. In order to investigate the property of the convective overshooting and to make it easy to apply turbulent convection models, we have analyzed the TCM developed by Li \& Yang \citep{li07} and obtained approximate and asymptotic solutions of the TCM in the overshooting region with $P_e \gg 1$. The main conclusions and corresponding discussions are listed as follows:

1. The overshooting region can be partitioned into three parts: a thin turbulent heat flux overshooting region, a power law dissipation region of turbulent kinetic energy, and a thermal dissipation area with rapidly decreasing $k$. The turbulent fluctuations $k$, $\overline{u_{r}'T'}$, and $\overline{T'T'}$ exponentially decrease in the overshooting region as Eqs.(34-36). The equilibrium value of the anisotropic degree $\omega_o$ and the exponential indices of the turbulent fluctuations are only determined by the parameters of the TCM. The decaying behaviors of the turbulent fluctuations are similar to Xiong \& Deng's results\citep{xio89,xio01}.

2. The peak of $\overline{T'T'}$ in the overshooting region is located at about $1H_k$ away from the convective boundary. In this distance, the modification of $\nabla$ caused by the overshooting is remarkable. An approximate profile of $\nabla$ comprises an adiabatic overshooting region with the length of $l_{ad}$ and an exponentially decreasing function, as described in Eq.(24) and (25). Beyond $1H_k$, the modification of $\nabla$ is ignorable and $\nabla \approx \nabla_R$. It should be noted that the result of $1H_k$ overshooting distance of turbulent heat transfer is independent of the parameters of TCM, so it may be a general property of the overshooting. Our result is similar to Marik \& Petrovay(2002) whose result shows that the length between the peak of $\overline{T'T'}$ and the convective boundary is about $1.2H_k$. Meakin \& Arnett(2010) simulated the turbulent convection of a $23M_{\odot}$ star, the data of the turbulent kinetic energy and the convective flux in the overshooting region being shown in Fig.8. It is found that the overshooting length of the convective flux $\overline{u_{r}'T'}$ is about $0.5 \sim 2H_k$ which is in agreement with our result.

3. The value of the turbulent kinetic energy at the convective boundary $k_C$ can be estimated by a method called \textsl{the maximum of diffusion}. The value of turbulent fluctuations in the overshooting region can be estimated by using the exponentially decreasing functions and the initial value $k_C$. This may significantly simplify the application of the TCM in calculations of the stellar structure and evolution.

There is a distinction between the non-local model of Zahn(1991) and our results, i.e. the
temperature gradient jumps from nearly adiabatic to radiative in Zahn's model but continuously changes in our results (see Fig.2). This is caused by the assumption in Zahn's model that the turbulent velocity and
temperature fluctuation are strongly correlated\citep{xio01}. In our results, the correlativity of turbulent velocity and
temperature fluctuation $R_{VT}=\frac{\overline{u_{r}'T'}}{\sqrt{2\omega k\overline{T'T'}}}$ quickly decreases to zero then turns to be negative near the convective boundary(see Fig.9), and the asymptotic solution shows that $R_{VT} \propto \sqrt{k}$ and exponentially decreases in the turbulent dissipation overshooting region. Our result is in agreement with three-dimension simulations such as Fig.6 in Singh et al.(1995) and Fig.15 in Meakin \& Arnett(2007).

\acknowledgments

We thank the anonymous referee for valuable comments
which help to improve the paper. And we thank C. A. Meakin for providing the numerical data of Fig.8. Fruitful discussions with J. Su, X. J. Lai and C. Y. Ding are highly appreciated. This work is co-sponsored by the National Natural Science Foundation of China through grant No.10673030 and No.10973035 and Science Foundation of Yunnan Observatory No.Y0ZX011009.

\appendix

\section{The localized TCM in convection zone.}

The localized TCM results from Eqs.(1-4) by ignoring the diffusion terms. It is a good approximate of the TCM in the convection zone\citep{li01}. We attempt to work out the solution in this appendix.

Some symbols are defined for conveniences: $U=\overline{u_{r}'T'}$, $V=\overline{T'T'}$, $W=\sqrt{k}$, $A=\frac{T}{H_P}(\nabla_R-\nabla_{ad})$,
$B=-\frac{\beta g_{r}}{T}$, $D=\frac{\lambda}{\rho C_P}$, $f=\frac{\nabla-\nabla_{ad}}{\nabla_R-\nabla_{ad}}$.

Ignoring the diffusion terms of Eqs.(1-4), we get the localized TCM:
\begin{eqnarray}
0=\frac{2}{3}\frac{W^3}{l} -2BU+2C_{k}( \omega -\frac{1}{3})\frac{W^3 }{l}
\end{eqnarray}%
\begin{eqnarray}
0=\frac{W^3}{l}-BU
\end{eqnarray}%
\begin{eqnarray}
0=-2\omega fAW^2-BV+C_{t}(1+P_e^{-1})\frac{WU}{l}
\end{eqnarray}%
\begin{eqnarray}
0=-2fAU+2C_{e}(1+P_e^{-1})\frac{WV}{l}
\end{eqnarray}%
\begin{eqnarray}
U=AD(1-f)
\end{eqnarray}%

Equation (A1) and (A2) show:
\begin{eqnarray}
\omega=\frac{2}{3C_{k}}+ \frac{1}{3}
\end{eqnarray}%

This is the equilibrium value $\omega_{cz}$ in convection zone.

Describing $W$, $V$, $U$ by $f$ and $P_e$($=\frac{lW}{D}$), we find:
\begin{eqnarray}
f=\frac{C_t C_e P_e^{-1}(1+P_e^{-1})^2}{C_t C_e P_e^{-1}(1+P_e^{-1})^2+2 C_e \omega (1+P_e^{-1}) +1}
\end{eqnarray}%

$W$, $V$ can be worked out as:
\begin{eqnarray}
W^3=ABDl(1-f)
\end{eqnarray}%
\begin{eqnarray}
V=\frac{AfW^2}{C_e B (1+P_e^{-1})}
\end{eqnarray}%

According to $P_e=\frac{lW}{D}$, Eq.(A8) and Eq.(A7), we get the equation of $P_e$:
\begin{eqnarray}
aP_e^4+(b+1)P_e^3+2bP_e^2+(b-at)P_e-t=0
\end{eqnarray}%
where $a=1+\frac{1}{2\omega C_e}$, $b=\frac{C_t}{2\omega}$, $t=\frac{ABl^4}{D^2}$. $f$ is determined by $f=1-\frac{P_e^3}{t}$ according to Eq.(A8).

Solving Eq.(A10), we can obtain all turbulent fluctuations of the localized TCM by using Eq.(A5), (A8), (A9) and (A11).

An important case is $t\gg1$, thus $P_e\gg1$ according to Eq.(A10). In that case, Eq.(A7) shows:
\begin{eqnarray}
f=\frac{C_e C_t P_e^{-1}}{2C_e \omega+1}\approx0
\end{eqnarray}%
which corresponds to the adiabatic convection.

Finally, we obtain the turbulent fluctuations according to Eq.(A8), (A5) \& (A9):
\begin{eqnarray}
W^3\approx ABDl
\end{eqnarray}%
\begin{eqnarray}
V\approx \frac{C_t}{2C_e \omega+1}\frac{AD}{Bl} W
\end{eqnarray}%
\begin{eqnarray}
U\approx AD
\end{eqnarray}%
and the correlativity of turbulent velocity and
temperature $R_{VT}$:
\begin{eqnarray}
R_{VT}=\frac{U}{\sqrt{2\omega W^2 V}}\approx\sqrt{\frac{2C_e \omega+1}{2C_t \omega}}
\end{eqnarray}%

\section{Details of deriving the asymptotic solution of the TCM in overshooting region.}

There are the details of obtaining the asymptotic solution of the TCM in overshooting region.

Some symbols are defined for conveniences: $U=\overline{u_{r}'T'}$, $V=\overline{T'T'}$, $W=\sqrt{k}$, $A=-\frac{T}{H_P}(\nabla-\nabla_{ad})\approx-\frac{T}{H_P}(\nabla_{R}-\nabla_{ad})$ (\textsl{Approximation III} is used),
$B=-\frac{\beta g_{r}}{T}$.

Applying the asymptotic property $\omega=\omega_o$ and \textsl{Approximations I, II \& III}, one can rewrite TCM as:
\begin{eqnarray}
0=(C_k-1)(\omega_o-\frac{1}{3})\frac{W^3}{l}-BU(1-\omega_o)
\end{eqnarray}%
\begin{eqnarray}
lC_s \omega_o \frac{\partial}{\partial r}(W\frac{\partial W^2}{\partial r})=\frac{W^3}{l}-BU
\end{eqnarray}%
\begin{eqnarray}
0=-BV+2A\omega_oW^2
\end{eqnarray}%
\begin{eqnarray}
lC_{e1}\omega_o\frac{\partial}{\partial r}(W\frac{\partial V}{\partial r})=AU+\frac{C_e}{l}WV
\end{eqnarray}%

Equation (B1) is equivalent to:
\begin{eqnarray}
U=\frac{(C_k-1)(\omega_o-\frac{1}{3})}{(1-\omega_o)}\frac{W^3}{Bl}
\end{eqnarray}%

Taking it into Eq.(B2), one gets the equation of $W$:
\begin{eqnarray}
\frac{\partial^2W^3}{\partial r^2}=\frac{3}{4C_s\omega_o l^2}[1-\frac{(C_k-1)(\omega_o-\frac{1}{3})}{(1-\omega_o)}]W^3
\end{eqnarray}%

Equation (B3) is equivalent to:
\begin{eqnarray}
V=\frac{2A\omega_o}{B}W^2
\end{eqnarray}%

According to Eq.(B4), (B5) and (B7), one gets another equation of $W$:
\begin{eqnarray}
\frac{\partial^2 W^3}{\partial r^2}=\frac{3}{4C_{e1}{\omega_o}^2l^2}[\frac{(C_k-1)(\omega_o-\frac{1}{3})}{(1-\omega_o)}+2C_e\omega_o]W^3
\end{eqnarray}%

Comparing Eq.(B6) with Eq.(B8), one finds:
\begin{eqnarray}
\frac{3}{4C_{e1}{\omega_o}^2l^2}[\frac{(C_k-1)(\omega_o-\frac{1}{3})}{(1-\omega_o)}+2C_e\omega_o]=\frac{3}{4C_s\omega_o l^2}[1-\frac{(C_k-1)(\omega_o-\frac{1}{3})}{(1-\omega_o)}]
\end{eqnarray}%

Therefore the equation of $\omega_o$ is:
\begin{eqnarray}
(2C_s C_e-C_{e1}C_k){\omega_o}^2+[\frac{1}{3}C_{e1}(C_k+2)-C_s(C_k+2C_e-1)]\omega_o+\frac{1}{3}C_s(C_k-1)=0
\end{eqnarray}

The asymptotic solution of $W$ is derived from Eq.(B6):
\begin{eqnarray}
W=W_0exp{\{\pm \frac{1}{2\alpha}\sqrt{\frac{1}{3C_s\omega_o}[1-\frac{(C_k-1)(\omega_o-\frac{1}{3})}{(1-\omega_o)}]}ln(\frac{P}{P_0})\}}
\end{eqnarray}%

$W$ takes the decreasing expression in the overshooting region: $'+'$ is adopted in the upward overshooting region and $'-'$ in the downward one.

\begin{figure}
\epsscale{1}
\plotone{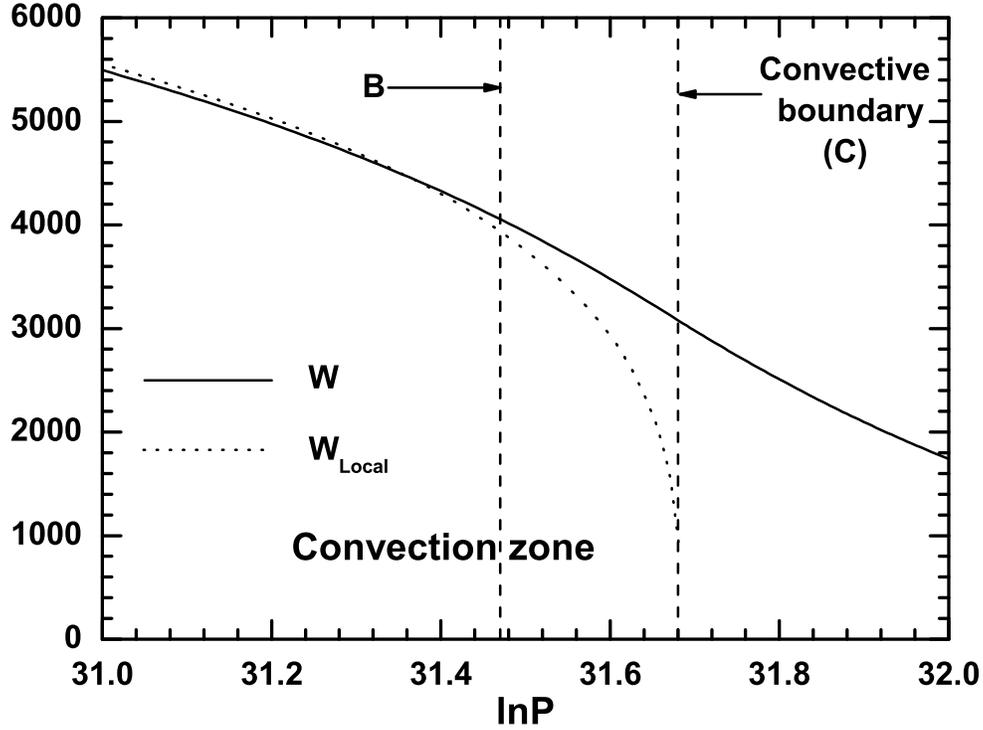}
\caption{Numerical results of $W=\sqrt{k}$, and $W_{Local}\approx \sqrt[3]{\frac{ \alpha\beta g\lambda(\nabla_{R}-\nabla_{ad})}{\rho c_{P}}}$ which is the solution of localized TCM (See Appendix A), for the solar model at present age. TCM parameters are: $\alpha=0.84$, $C_k=2.5$, $C_s=0.1$, $C_{e1}=0$, $C_e=0.2$, $C_t=7.0$, and $C_{t1}=0.01$, but $C_t$, $C_{t1}$ and $C_{e1}$ are insensitive to the results. Point C is the boundary of the convection zone, the location of point B is calculated by using Eq.(22).}
\label{Fig.1}
\end{figure}

\begin{figure}
\epsscale{1}
\plotone{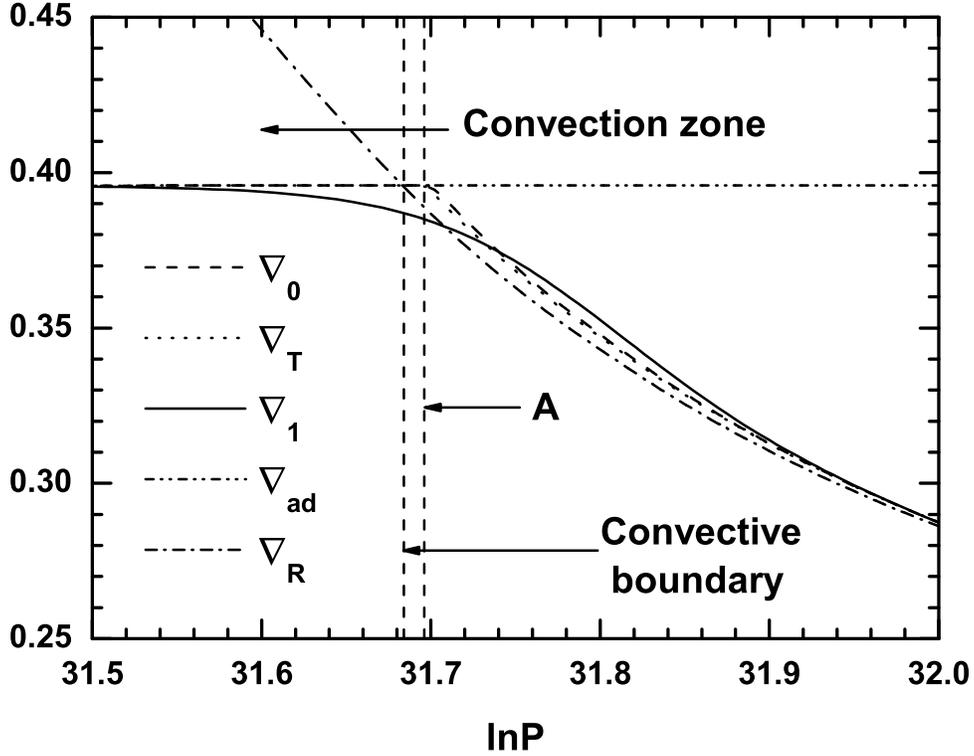}
\caption{Numerical results of temperature gradient near the convective boundary in both $C_{e1}=0$ and $C_{e1}\neq0$, $\nabla_0$ being the temperature gradient of the model with $C_{e1}=0$, and $\nabla_1$ corresponding to $C_{e1}=0.01$. Dotted line $\nabla_T$, which is almost identical to $\nabla_0$, is theoretical solution of the temperature gradient with $C_{e1}=0$. The stellar model and other TCM parameters are the same as Fig.1. Point A is the boundary of the adiabatic overshooting region. Our theoretical result shows $l_{ad}\approx0.013H_P$ in those TCM parameters set, the numerical calculation being $0.015H_P$.}
\label{Fig.2}
\end{figure}

\begin{figure}
\epsscale{1}
\plotone{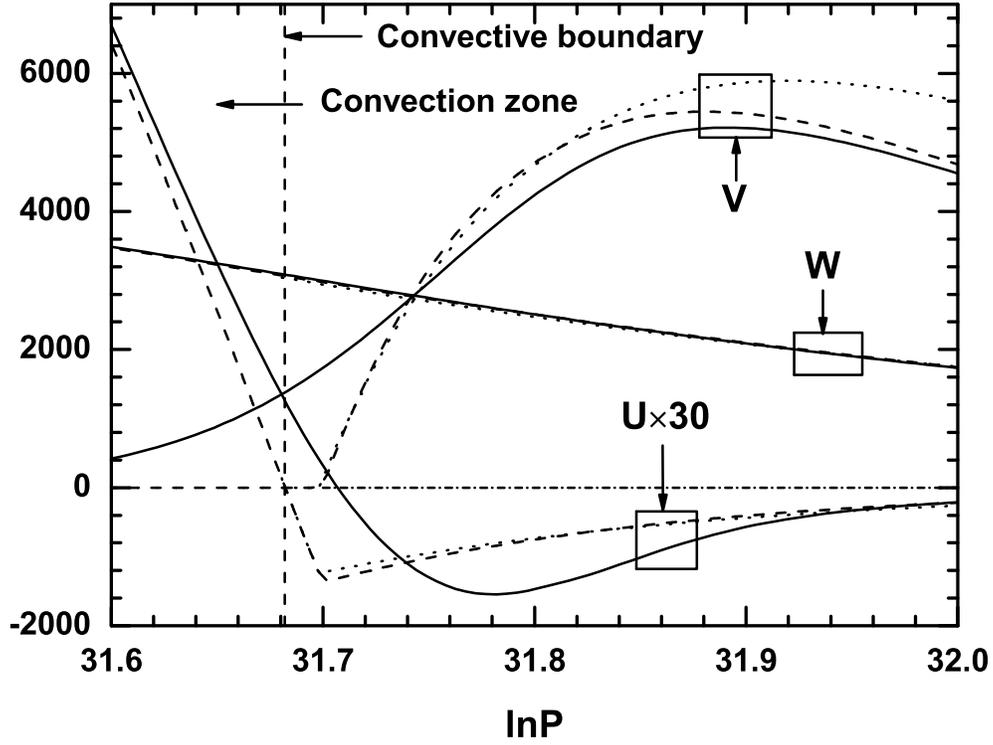}
\caption{Numerical results of $\overline{T'T'}$, $\overline{u_r'T'}$, $k$ near the convective boundary in both $C_{e1}=0$ and $C_{e1}\neq0$, where $U=\overline{u_r'T'}$, $W=\sqrt{k}$, $V=\overline{T'T'}$. Dashed lines correspond to $C_{e1}=0$, solid lines to $C_{e1}=0.01$. Dotted lines are the theoretical solutions with $C_{e1}=0$. The stellar model and other TCM parameters are the same as Fig.1.}
\label{Fig.3}
\end{figure}

\begin{figure}
\epsscale{1}
\plotone{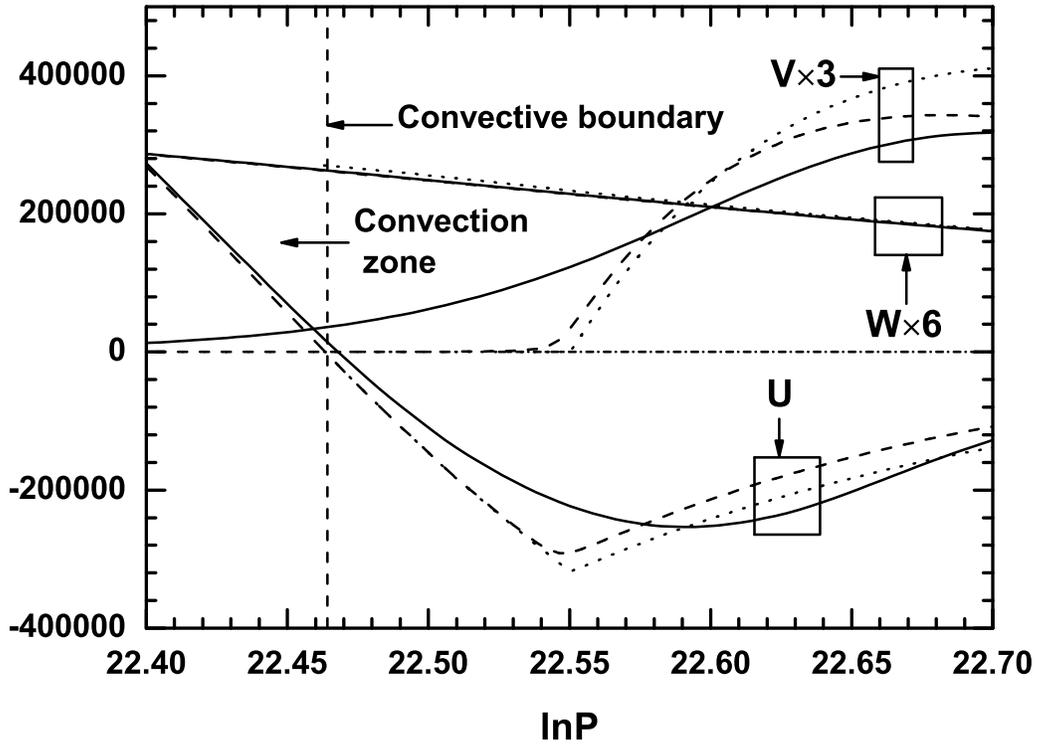}
\caption{Numerical results of $\overline{T'T'}$, $\overline{u_r'T'}$, $k$ near the convective boundary in both $C_{e1}=0$ and $C_{e1}\neq0$, where $U=\overline{u_r'T'}$, $W=\sqrt{k}$, $V=\overline{T'T'}$. Dashed lines correspond to $C_{e1}=0$, solid lines to $C_{e1}=0.01$. Dotted lines are the theoretical solutions with $C_{e1}=0$. The stellar model is a $7M_\odot$ star model at the top of RGB phase. Others TCM parameters are: $\alpha=1.0$, $C_k=2.2$, $C_s=0.1$, $C_e=1.0$, and $C_t=4.0$, $C_{t1}=0.01$.}
\label{Fig.4}
\end{figure}

\begin{figure}
\epsscale{1}
\plotone{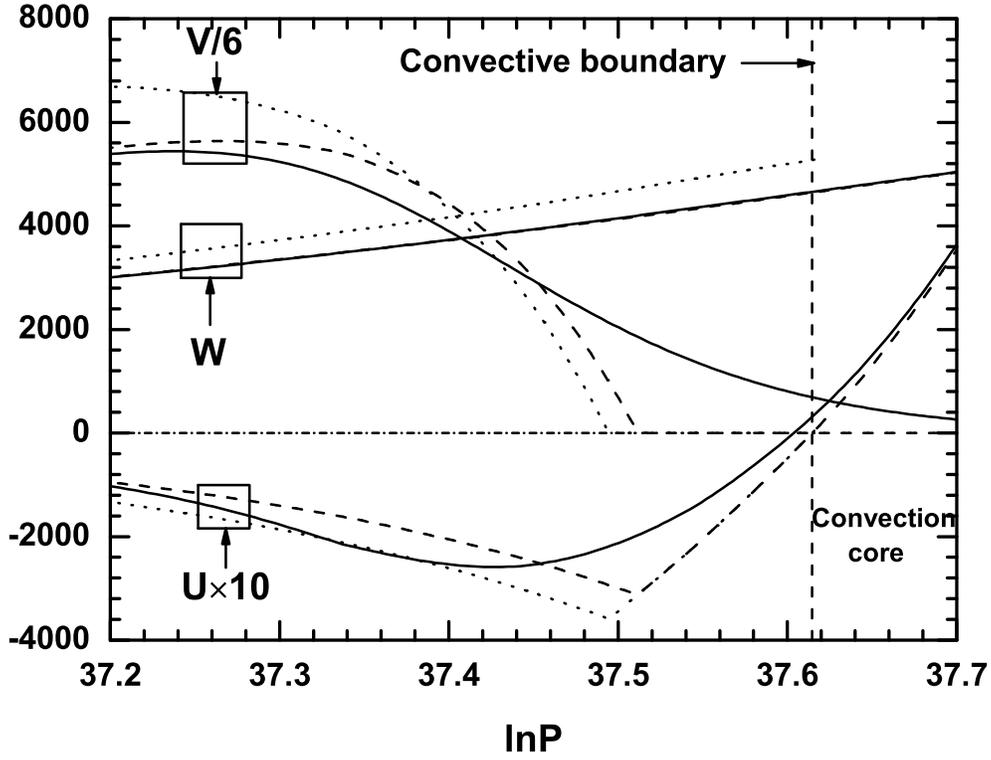}
\caption{Numerical results of $\overline{T'T'}$, $\overline{u_r'T'}$, $k$ near the boundary of the convective core in both $C_{e1}=0$ and $C_{e1}\neq0$, where $U=\overline{u_r'T'}$, $W=\sqrt{k}$, $V=\overline{T'T'}$. Dashed lines  correspond to  $C_{e1}=0$, solid lines to $C_{e1}=0.01$. Dotted lines are the theoretical solutions with $C_{e1}=0$. The stellar model is an early main sequence model of a $3M_\odot$ star. Others TCM parameters are: $\alpha=1.0$, $C_k=2.1$, $C_s=0.2$, $C_e=0.5$, and $C_t=3.0$, $C_{t1}=0.01$.}
\label{Fig.5}
\end{figure}

\begin{figure}
\epsscale{1}
\plotone{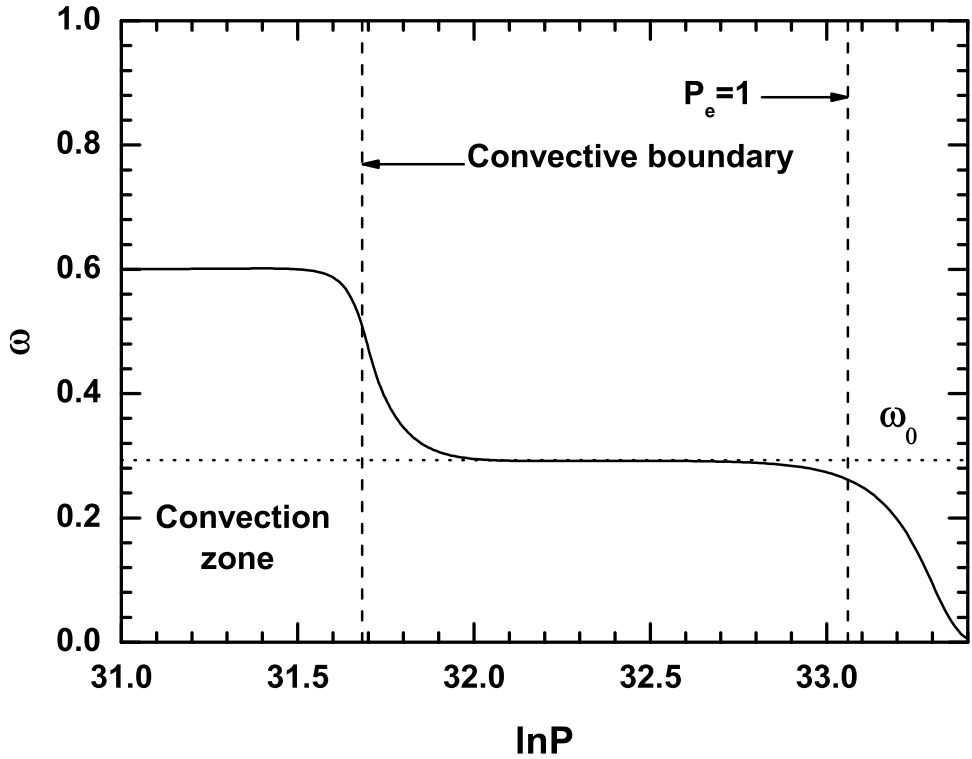}
\caption{Numerical result of the structure of $\omega$ in overshooting region. The stellar model is the solar model at present age. $C_{e1}=0.01$. The others TCM parameters are the same as Fig.1, except $\alpha =0.2$ in order to enlarge $\theta$ to show the thermal dissipation region in which $P_e\ll 1$. With those parameters, the equilibrium value in convection zone is $\omega_{cz}=0.6$, and the equilibrium value in overshooting region is $\omega_o=0.293$ which denoted as the dotted line.}
\label{Fig.6}
\end{figure}

\begin{figure}
\epsscale{1}
\plotone{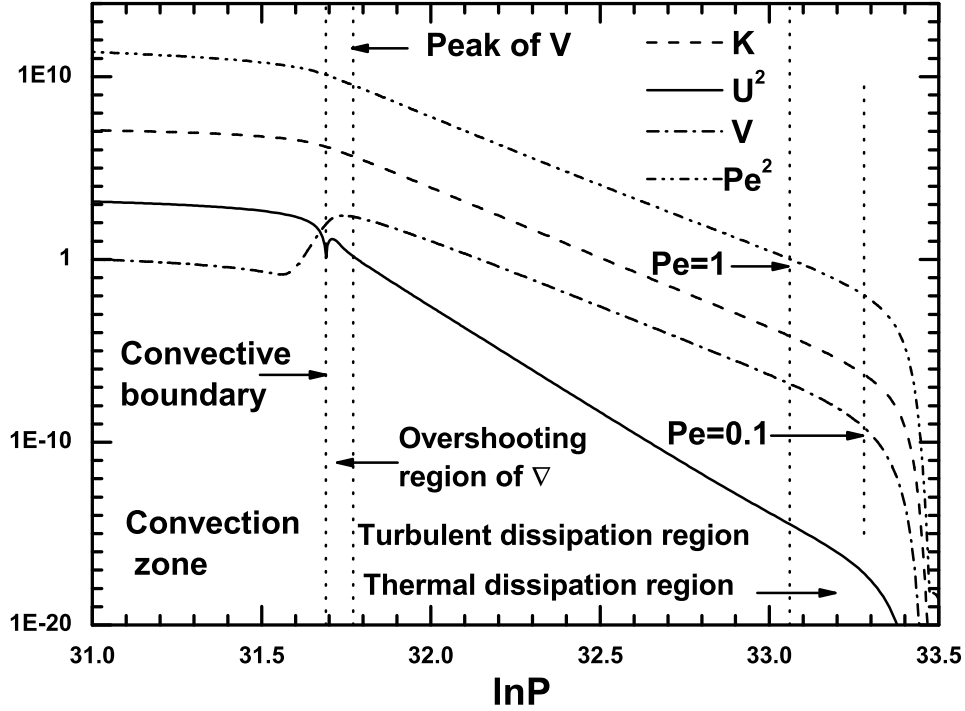}
\caption{The structure of the overshooting region. $K=k$, $U=\overline{u_r'T'}$, $V=\overline{T'T'}$. The stellar model is the solar model at present age. $C_{e1}=0.01$, the others TCM parameters are the same as Fig.1, except $\alpha =0.2$. With those parameters, in the turbulent dissipation region with $P_e\gg 1$, theoretical result shows $\theta=17.5$ vs the numerical result $17.6$, theoretical result of exponential decreasing index of $U^2$ being $26.3$ vs the numerical result about $25.6$. $K$ is almost parallel to $V$, which is in consistent with the asymptotic solution. In the thermal dissipation region with $P_e\ll 1$, turbulent motion vanishes.}
\label{Fig.7}
\end{figure}

\begin{figure}
\epsscale{1}
\plotone{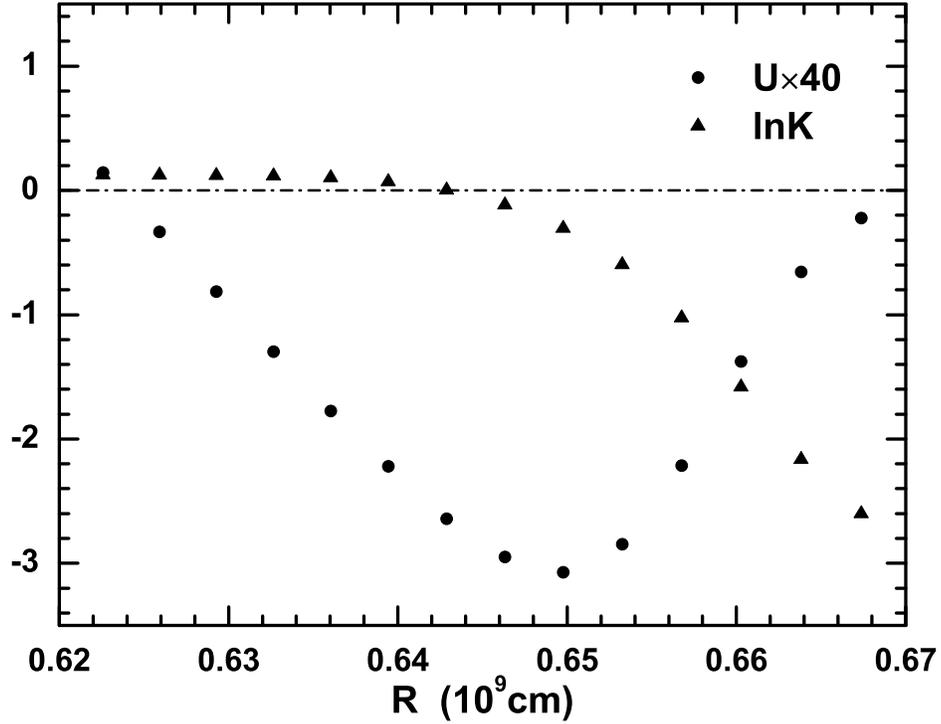}
\caption{Numerical data of Meakin \& Arnett (2010)'s results. The data of model 'h1' in their paper are plotted, where $U=F_C=\rho C_P\overline{u_r'T'}$. Only the downward overshooting region is shown. The distance from the convective boundary (where $\overline{u_{r}'T'}=0$, about $R=0.62\times10^9cm$) to the right part of the valley of $\overline{u_{r}'T'}$ is about $0.5 \sim 2H_k$.}
\label{Fig.8}
\end{figure}

\begin{figure}
\epsscale{1}
\plotone{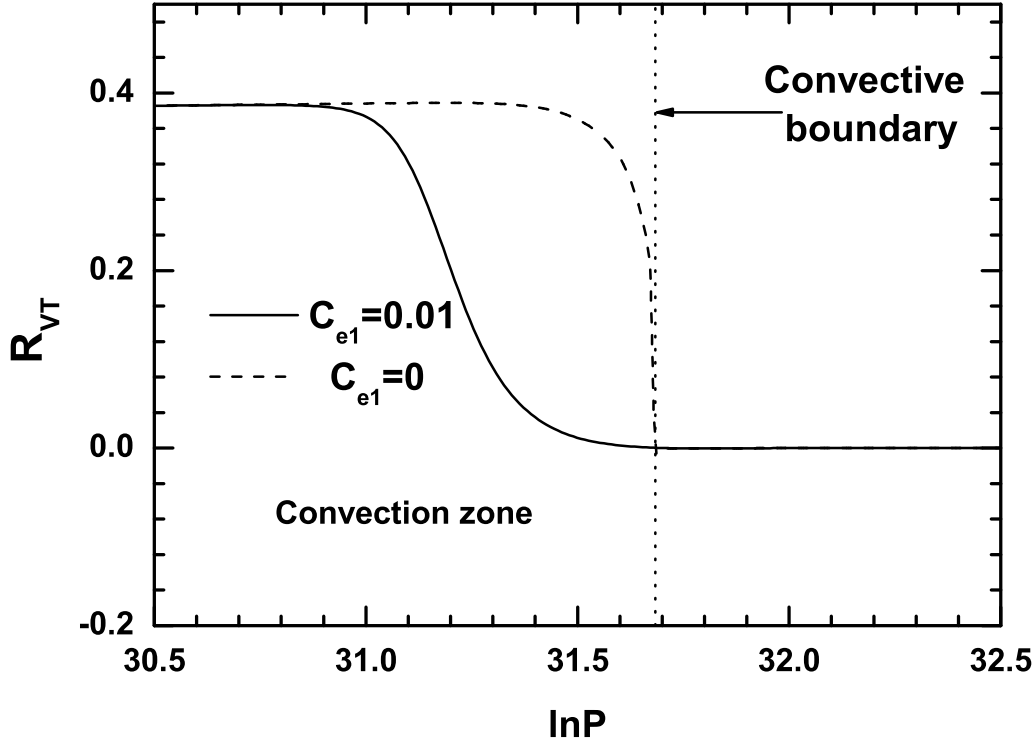}
\caption{Numerical results of the correlativity of turbulent velocity and
temperature $R_{VT}$. The stellar model is the solar model at present age. Other TCM parameters are the same as Fig.1. $R_{VT}$ rapidly decreases to zero in the overshooting region. In the convection zone near the convective boundary, the diffusion significantly enlarges $\overline{T'T'}$ when $C_{e1}\neq0$ (see Fig.3), and then $R_{VT}$ is very small. In the interior of convection zone, localized TCM shows the equilibrium value of $R_{VT}$ is $R_{VT,cz}=\sqrt{\frac{2\omega_{cz} C_e +1}{2\omega_{cz} C_t}}$ (see Appendix A). The TCM parameters show $R_{VT,cz}=0.384$.} \label{Fig.9}
\end{figure}

\end{document}